# Higher odd-order nonlinear Hall effect in magnetic topological insulator Mn(Bi$_{1-x}$Sb$_x$)$_2$Te$_4$

Xiubing Li[1,†], Zheng Dai[1,†], Shuai Zhang[1,2,3,*], Heng Zhang[1], Congcong Li[1], Boyuan Wei[1], Fengyi Guo[1], Chunfeng Li[1], Fucong Fei[4], Minhao Zhang[1,2,3], Xuefeng Wang[5], Huaiqiang Wang[6,*], and Fengqi Song[1,2,3,7,*]

[1] National Laboratory of Solid State Microstructures, Collaborative Innovation Center of Advanced Microstructures, Jiangsu Physical Science Research Center, and School of Physics, Nanjing University, Nanjing 210093, China

[2] Institute of Atom Manufacturing, Nanjing University, Nanjing 211800, China

[3] Nanjing Institute of Atomic Scale Manufacturing, Nanjing 211800, China

[4] School of Materials Science and Intelligent Engineering, Nanjing University, Suzhou 215163, China

[5] Jiangsu Provincial Key Laboratory of Advanced Photonic and Electronic Materials, State Key Laboratory of Spintronics Devices and Technologies, School of Electronic Science and Engineering, Nanjing University, Nanjing 210093, China

[6] Center for Quantum Transport and Thermal Energy Science, Institute of Physics Frontiers and Interdisciplinary Sciences, School of Physics and Technology, Nanjing Normal University, Nanjing 210023, China

[7] Suzhou Laboratory, Suzhou 215125, China

[†]These authors contributed equally.

[*]Corresponding authors. S.Z. (szhang@nju.edu.cn), H.W. (hqwang@njnu.edu.cn), F.S. (songfengqi@nju.edu.cn)

**Abstract**

The nonlinear Hall effect is a new member of the Hall effect family, which attracts intense research interests, and it is closely related to the quantum geometry of quantum materials. The previous studies primarily concentrate on the second-order and third-order nonlinear Hall effect. However, the experimental study of higher-order nonlinear Hall effect is scarce at present. In this work, we report the observations of the higher odd-order (third-, fifth-, seventh-order) nonlinear Hall effect in magnetic topological insulator $Mn(Bi_{1-x}Sb_x)_2Te_4$ thin flakes. The higher odd-order nonlinear Hall voltage exhibits a twofold angular dependence and exists only below the Néel temperature. It reaches its maximum near the charge neutral point and decays exponentially as the order of the nonlinear Hall effect increases. Furthermore, such higher odd-order nonlinear Hall effect is observed in both odd- and even-layer samples with comparable magnitudes. Theoretical analysis indicates that the higher odd-order nonlinear Hall effect responses may arise from the Berry curvature multipoles. Our work paves the way for the study of the higher-order nonlinear transport phenomena.
.

**Introduction**

The Hall effect has been a prominent topic in the field of condensed matter physics since its discovery, leading to a series of subsequent fundamental investigations and practical applications. It specifically refers to the phenomenon where a transverse voltage perpendicular to the longitudinal current is generated when an external magnetic field or intrinsic magnetism breaks the time-reversal symmetry (TRS)[1,2]. Recently, a second-order nonlinear Hall effect (NLHE) has been observed in non-centrosymmetric quantum materials without breaking TRS[3-5], manifested by a second-harmonic transverse Hall voltage with quadratic dependence on the longitudinal driving current. As a new member of the Hall effect family, it arises from either intrinsic Berry curvature dipole (BCD)[3-9], quantum metric dipole[10,11] or extrinsic scatterings[12-17].

NLHE has attracted significant research interests since its discovery[18,19]. Recently, the third-order NLHE has been observed[20-27], showing cubic dependence on the longitudinal current. This response is primarily attributed to quantum geometry quadrupole and some other mechanisms. Furthermore, the room-temperature NLHE has been realized[28-34]. And the radiofrequency rectification based on the NLHE has also been demonstrated[28,30,34-36], showcasing its distinctive advantages for device applications. These advances in NLHE have significantly propelled the advancement of the field. Current researches on NLHE concentrate predominantly on second- and third-order phenomena[37-41]. Although research on NLHE has attracted growing attention in terms of higher-order effects[18,42-46], and corresponding theoretical work has been conducted, systematic experimental exploration is still lacking.

Magnetic topological insulator MnBi$_2$Te$_4$ (MBT) systems[47,48] establish an exceptional platform for studying higher-order NLHE phenomena, leveraging their unique symmetry properties, topological surface states, and magnetic configurations. In MnBi$_2$Te$_4$, the second-order[10,11] and third-order[26] NLHE mediated by the quantum geometry dipole and quadrupole have been studied below the Néel temperature. And Sb-doped MnBi$_4$Te$_7$ exhibits a large second-order NLHE persisting above its magnetic transition temperature[49].

In this work, we study the higher-order NLHE in magnetic topological insulator Mn(Bi$_{1-x}$Sb$_x$)$_2$Te$_4$ ($x \approx 0.3$) thin flakes. The higher odd-order NLHE is observed, which nonlinear Hall voltage satisfies a current dependence of $V_{xy}^{(2n+1)\omega} \propto (I^\omega)^{2n+1}$ and exhibits a twofold angular dependence. It is exclusively present below the Néel temperature and can be effectively modulated by the gate voltage, achieving its maximum value near the charge neutral point. Furthermore, it decays exponentially with the NLHE order increasing. Theoretical analysis indicates that the higher odd-order NLHE here originates from Berry curvature multipoles (BCMs). Besides, the observed higher odd-order NLHE exhibits almost thickness-independent response behaviors, which could be attributed to mismatched out-of-plane surface magnetic moments on the top and bottom layers. This work enriches the research on higher-order NLHE experimentally, and advances fundamental understanding of nonlinear electric transport response.

## Results

**Basic characteristics of 5SL Mn(Bi$_{1-x}$Sb$_x$)$_2$Te$_4$**

The Mn(Bi$_{1-x}$Sb$_x$)$_2$Te$_4$ (Sb-MBT) is composed by stacking of Te-X-Te-Mn-Te-X-Te (X = Bi/Sb) septuple layers (SLs) along the *c* axis through van der Waals interactions[50,51]. It exhibits A-type antiferromagnetism (AFM), characterized by intralayer ferromagnetic (FM) coupling and interlayer AFM coupling[52]. Doping with Sb atoms enables the modulation of the Fermi level and topological properties [50,53]. At $x = 0.3$, the Fermi level lies within the band gap and the topological properties remain unchanged, facilitating studies of nonlinear transport phenomena.

We fabricated disc-shaped Sb-MBT device with 12 electrodes (Fig. 1a) by electron beam lithography and electron beam evaporation. Hexagonal boron nitride (*h*-BN) and graphite were sequentially dry-transferred onto the Sb-MBT thin flake. The graphite/*h*-BN stack functions as the top gate ($V_{tg}$). For transport measurement, a driving a.c. current $I^\omega$ (frequency $\omega$ = 23 Hz) is applied between a pair of the 12 electrodes at an angle $\theta$ (with $\theta = 0°$ defined independently of the lattice orientation), and the longitudinal and Hall voltages are measured simultaneously.

An optical image of 5SL Sb-MBT device after transferring graphite/*h*-BN stack is shown in Fig. 1b. Figure 1c displays the temperature (*T*) dependence of longitudinal resistance ($R_{xx}$), exhibiting the typical transport characteristics of thin Sb-MBT samples. The magnetic ordering temperature (Néel temperature $T_N$) is identified at approximately 24 K[53-55]. Figures 1d,e show the magnetic field (**B**) dependence of $R_{xx}$ and $R_{xy}$. An obvious anomalous Hall effect is observed, characterized by a coercive field of about 0.2 T. The top gate can effectively modulate the carrier density, enabling access to the charge neutral point (CNP), which occurs at at $V_{tg}$ = 1 V, as clearly demonstrated in Fig. 1f.

**Observation of higher odd-order nonlinear Hall effect in Mn(Bi$_{1-x}$Sb$_x$)$_2$Te$_4$**

In the 5SL Sb-MBT device, we explore the odd-order nonlinear Hall voltages $V_{xy}^{(2n+1)\omega}$ (Fig. 2). The second-order nonlinear Hall response ($V_{xy}^{2\omega}$) is negligibly small compared to the odd-order responses (see Supplementary Note 1 for the detailed discussions of $V_{xy}^{2\omega}$). We first characterize the third-order nonlinear Hall voltage $V_{xy}^{3w}$. At 5 K and zero magnetic field, the $V_{xy}^{3w}$ signal is clearly detected at every angle $\theta$, exhibiting a cubic dependence on the applied current $V_{xy}^{3\omega} \propto (I^{\omega})^3$, as shown in Fig. 2a. The maximum amplitude of $|V_{xy}^{3\omega}|$ reaches 25 μV under an applied current $I^{\omega}$ = 5 μA at $\theta$ = 120° or 300°.

Interestingly, subsequent measurements have revealed the presence of higher odd-order nonlinear Hall voltages. As shown in Figs. 2b,c, both fifth-order ($V_{xy}^{5w}$) and seventh-order ($V_{xy}^{7w}$) nonlinear Hall voltages are observed, maintaining the expected current scaling $V_{xy}^{5\omega} \propto (I^{\omega})^5$ and $V_{xy}^{7\omega} \propto (I^{\omega})^7$. Under the same condition as $V_{xy}^{3\omega}$, the maximum values of $|V_{xy}^{5w}|$ and $|V_{xy}^{7w}|$ can reach 3.5 μV and 0.7 μV, respectively. Notably, at lower temperature, the additional higher odd-order nonlinear Hall voltages $V_{xy}^{9\omega}$ and $V_{xy}^{11\omega}$ are also observed (Supplementary Fig. 2). These observations demonstrate the presence of higher odd-order NLHE in Sb-MBT.

Figure 2d illustrates the angular dependence of nonlinear Hall voltages $V_{xy}^{(2n+1)\omega}$ at 5 K under $I^{\omega}$ = 5 µA. They all exhibit pronounced twofold symmetry with 180° periodicity. When the directions of current and Hall voltage are reversed simultaneously, the curves of $V_{xy}^{(2n+1)\omega}$ overlap. Both $V_{xx}^{1\omega}$ and $V_{xy}^{1\omega}$ exhibit a linear relationship with $I^{\omega}$, as shown in Fig. 2e. It is noteworthy that the angle-dependent $V_{xx}^{1\omega}$ also demonstrates a twofold symmetry with a periodicity of 180° (Supplementary Fig. 5). This twofold symmetry may be related to the doping of Sb atoms[56-58].

To further investigate the higher odd order NLHE here, it is imperative to first rule out contributions from artificial effects, such as capacitance coupling effect, contact junction effect, Joule heating effect. We have excluded the impact of these artificial effects on the measurement of the higher odd-order NLHE (Supplementary Note 3).

**Gate and temperature-dependent higher odd-order nonlinear Hall effect in Mn(Bi$_{1-x}$Sb$_x$)$_2$Te$_4$**

To better understand the properties of higher odd-order NLHE, we further study its gate dependence. The higher odd-order nonlinear Hall voltage $V_{xy}^{(2n+1)\omega}$ as the function of top gate voltages at 5 K and $I^{\omega}$ = 5 µA is shown in Fig. 3a, with the angle $\theta$ fixed at 330°. The $V_{xy}^{(2n+1)\omega}$ exhibits an obvious peak, reaching its maximum value near the CNP, as indicated by the dashed line in Fig. 3a. When $V_{tg}$ is distant from both sides of the CNP, the strength of $V_{xy}^{(2n+1)\omega}$ decreases rapidly. Mapping

$V_{xy}^{(2n+1)\omega}$ across various gate voltages and temperatures reveals that there is a peak near the CNP under all measured temperatures (Supplementary Fig. 4).

Figure 3b depicts the temperature dependence of higher odd-order NLHE. It can be seen that the $V_{xy}^{(2n+1)\omega}$ only exists below $T_N$. But the onset temperature ($T_O$) of the $V_{xy}^{(2n+1)\omega}$ is different. The $T_O$ of $V_{xy}^{3\omega}$ is about 20 K, while the $T_O$ of $V_{xy}^{5\omega}$ and $V_{xy}^{7\omega}$ is near 10 K and 7 K, respectively. As the temperature decreases, each nonlinear Hall signal appears and subsequently exhibits accelerated growth. At 1.8 K under a drive current $I^{\omega} = 5$ μA, the magnitude of the $|V_{xy}^{3\omega}|$ reaches 28.5 μV, and it is 7.5 μV for $|V_{xy}^{5\omega}|$, 2.7 μV for $|V_{xy}^{7\omega}|$. Since the intensity of higher odd-order NLHE decreases with increasing order, a lower temperature is necessary to observe the higher-order NLHE. Generally, the generation efficiency of NLHE[49,59] is quantitatively characterized by $V_{xy}^{(2n+1)\omega}/(V_{xx}^{1\omega})^{2n+1}$. It is observed that the $V_{xy}^{(2n+1)\omega}/(V_{xx}^{1\omega})^{2n+1}$ shares the same temperature-dependent behavior as $V_{xy}^{(2n+1)\omega}$ (Supplementary Fig. 3), which are 12.8 V$^{-2}$ for $V_{xy}^{3\omega}$, 2.2×10$^4$ V$^{-4}$ for $V_{xy}^{5\omega}$, and 4.6×10$^7$ V$^{-6}$ for $V_{xy}^{7\omega}$ at 1.8 K.

Furthermore, we can also see that the odd-order nonlinear Hall voltage $V_{xy}^{(2n+1)\omega}$ decreases remarkably with the nonlinear order $N$ increasing, where $N = 2n +1$. As shown in Fig. 3c, the $\log V_{xy}^{(2n+1)\omega}$ conforms a linear relationship with the nonlinear order $N$, which means that the $V_{xy}^{(2n+1)\omega}$ decreases exponentially with increasing $N$. Higher-order harmonics have been extensively studied in the field of nonlinear optics[60-62]. But the exponential attenuation behavior observed here differs from that in nonlinear optics, and the physical origin demands further study.

The higher odd-order NLHE can be reproduced in other odd-layer Sb-MBT

samples (Supplementary Note 2). All the characteristic behaviors (including periodicity, gate tunability, temperature dependence, and exponential attenuation) are successfully reproduced in these samples. Besides, the higher odd-order NLHE is also observed in 6SL Sb-MBT flake (Supplementary Figs. 17-19).

**Discussions**

Now we will show that the observed higher odd-order NLHE in the Sb-MBT thin flake most probably originates from BCMs[42]. As a magnetic topological insulator, the low energy physics of Sb-MBT flake near the CNP is dominated by the Dirac-cone surface states[51,63-67], which can be simply described by the effective Hamiltonian $H(\mathbf{k}) = \pm v(k_x \sigma_y - k_y \sigma_x) + m\sigma_z$. Here, $+(-)$ represents the top (bottom) surface, $v$ is the Fermi velocity, $\sigma_i$ ($i = x, y, z$) are Pauli matrices acting in the spin subspaces, and $m$ is the TRS-breaking Dirac mass induced by the out-of-plane surface magnetic moments. For the 5SL Sb-MBT flake (Fig. 4a), the magnetic moments of the top and bottom SLs are aligned in the same direction, leading to the same mass term and identical Berry curvatures for the two surface states. The Berry curvatures from surface states only have out-of-plane ($z$) components, given by $\Omega = \pm mv^2/(2d^3)$, where $+(-)$ denotes the conduction (valence) band with $d = \sqrt{m^2 + v^2 k_x^2 + v^2 k_y^2}$. The integration of these Berry curvatures over occupied states for each surface state contributes to the conventional first-order anomalous Hall effect[1], as shown in Fig. 4e.

As for the higher-order NLHE response, it has been shown that the ($N$+1)-th order NLHE is contributed by the $N$-th moment of Berry curvature[42] (Supplementary Note 6), which possesses $N$+1 independent components in two-dimensional systems: $\int_k f_0 (\partial_x)^l (\partial_y)^{N-l} \Omega$ with $l = 0, 1, \ldots, N$. Note that since the average crystal structure of the odd-SL Sb-MBT retains the inversion symmetry due to the random homogeneous doping in our samples, all odd-order moments of the Berry curvature vanish, thus suppressing even-order NLHE responses. As for the odd-order NLHE responses, the Figs. 4b-d present typical $k$-resolved distributions for the dominant components of Berry curvature quadrupole ($q_{xx} = \partial_x^2 \Omega$), octapole ($o_{xxxx} = \partial_x^4 \Omega$), and dodecapole ($t_{xxxxxx} = \partial_x^6 \Omega$), corresponding to the 2nd-, 4th-, and 6th-order moment of Berry curvature, respectively (Supplementary Note 6). The integrated Berry curvature quadrupole, octapole, and, dodecapole, as a function of the Fermi level at $T$ = 5 K are presented in Figs. 4f-h, respectively, where all of them exhibit peaks near the CNP, which are consistent with experimentally observed third-order, fifth-order, and seventh-order NLHE responses.

In addition, we clarify key aspects concerning the comparison between the theoretical calculations and experiments. First, above the Néel temperature, the Dirac mass term vanishes, thus eliminating Berry curvatures and their multipoles, in consistence with the observed disappearance of higher-order NLHE above the Néel temperature. Second, although the Dirac model is assumed to be isotropic, we notice that the Sb-doping process introduces an emergent $C_{2z}$ rotational symmetry of the average crystal structure, resulting in the observed two-fold symmetric behaviors of all the higher-order responses. Third, the NLHE response, as a Fermi surface property, should vanish within the surface state energy gap. However, in real experiments, there may exist a slight energy shift (~meV, comparable to the Dirac mass) between the top and bottom Dirac surface states due to different electrostatic potentials. As a result, the total NLHE responses contributed by the two misaligned surface states remain nonzero near the CNP. Fourth, to explain the observed nearly thickness-independent behavior of the NLHE, where the magnitudes are comparable in odd- and even-layer samples, we performed layer-resolved calculations based on the multi-Dirac-fermion mode[63,65] (Supplementary Note 7). Our results demonstrate that intrinsic higher odd-order NLHE near the CNP originate predominantly from surface states, with bulk states playing a negligible role. We also show that similar NLHE responses can emerge in both odd- and even-layer samples, originating from a mismatch between the out-of-plane magnetic moments of the top and bottom layers due to suppressed moments in the top SL[68] (Supplementary Fig. 24).

Furthermore, although the third-order NLHE has been reported in thick MBT[26], an external magnetic field is required for this observation. In contrast, our work demonstrates the observation of third-order NLHE, and even higher-order NLHE, without relying on an external magnetic field. And the BCMs of Sb-MBT systems are also not studied in previous theoretical work[42].

In summary, we report the higher odd-order NLHE in the magnetic topological insulator Mn(Bi$_{1-x}$Sb$_x$)$_2$Te$_4$ thin flakes. Various transport characteristics, dependency relationships, and the exponential attenuation law are systematically analyzed and explored. We surmise this higher odd-order nonlinear response originates from the BCMs. Our work has enhanced a more comprehensive understanding of nonlinear transport response.

**Figure Captions**

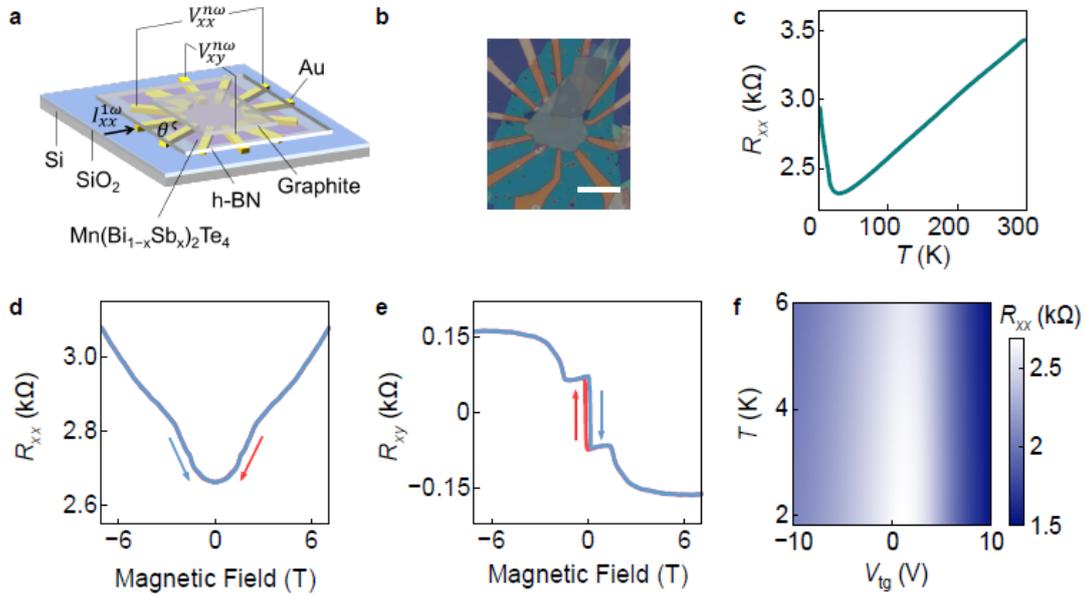

**Fig. 1 | Basic characteristics of 5SL Mn(Bi$_{1-x}$Sb$_x$)$_2$Te$_4$.**

**a,** Schematic of the disc-shaped device. The graphite/*h*-BN layer works as top gate. **b,** Optical image of the 5SL Mn(Bi$_{1-x}$Sb$_x$)$_2$Te$_4$ device. The white scale bar is 20 μm. **c,** Temperature-dependent longitudinal resistance. **d,** Magnetic field dependence of longitudinal resistance at 1.8 K. **e,** Magnetic field dependence of Hall resistance at 1.8 K. **f,** Top gate voltage ($V_{tg}$) dependence of longitudinal resistance at different temperatures. The charge neutral point is located at $V_{tg}$ = 1 V.

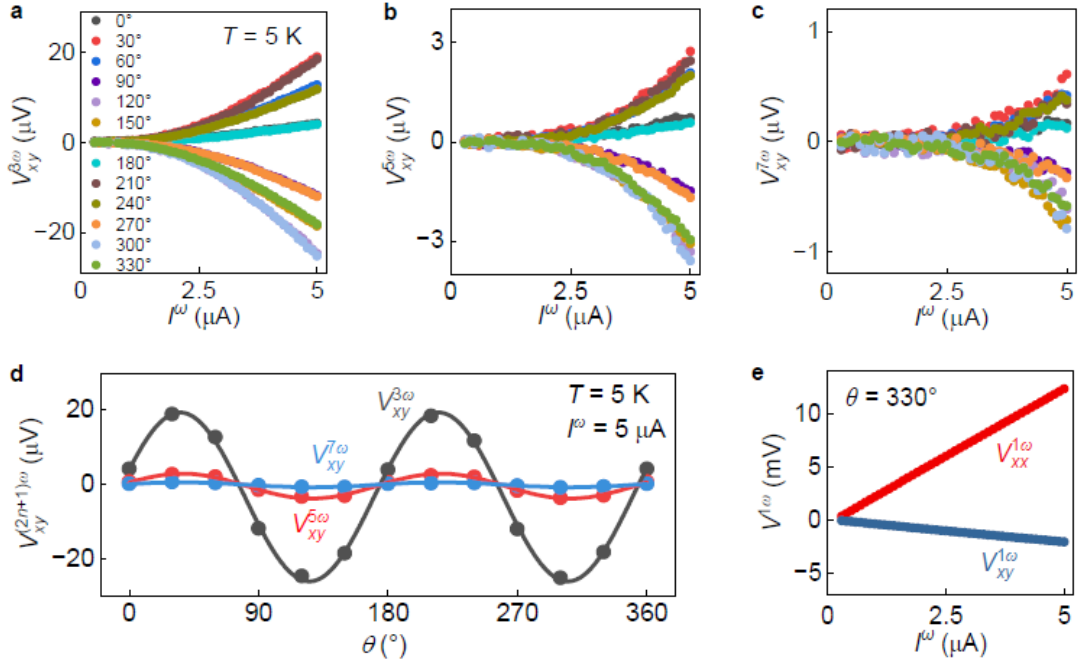

**Fig. 2 | Observation of higher odd-order nonlinear Hall effect in 5SL Mn(Bi$_{1-x}$Sb$_x$)$_2$Te$_4$.**

**a,** The current $I^\omega$ ($\omega$ = 23 Hz) dependence of the third-order nonlinear Hall voltage $V_{xy}^{3\omega}$ at various angles under 5 K. **b,** The fifth-order nonlinear Hall voltage $V_{xy}^{5\omega}$. **c,** The seventh-order nonlinear Hall voltage $V_{xy}^{7\omega}$. The relationship of $V_{xy}^{(2n+1)\omega} \propto (I^\omega)^{2n+1}$ is satisfied. **d,** The angle-dependent $V_{xy}^{(2n+1)\omega}$ shows a periodicity of 180° at $I^\omega$ = 5 μA. **e,** Linear current dependence of $V_{xx}^{1\omega}$ and $V_{xy}^{1\omega}$ at $T$ = 5 K and $\theta$ = 330°.

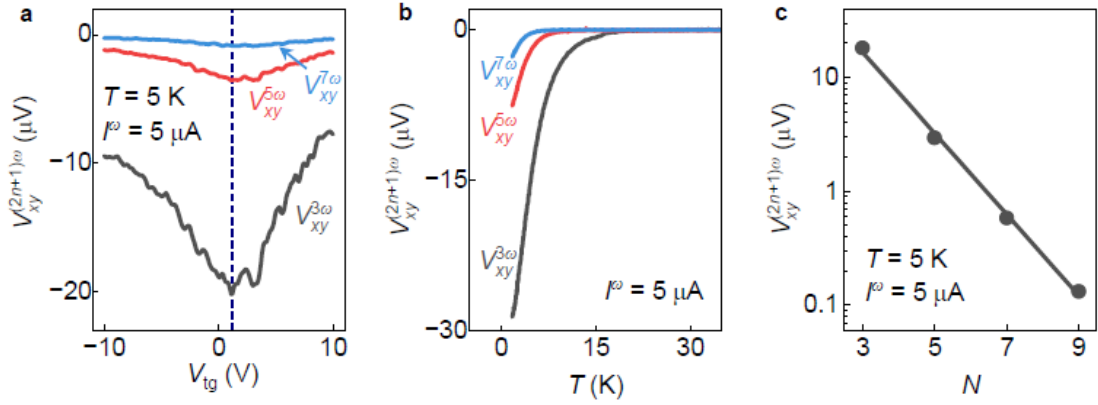

**Fig. 3 | Gate- and temperature-dependent higher odd-order nonlinear Hall effect in 5SL Mn(Bi$_{1-x}$Sb$_x$)$_2$Te$_4$.**

**a,** $V_{tg}$-dependent higher odd-order nonlinear Hall voltage $V_{xy}^{(2n+1)\omega}$ at $T = 5$ K, $I^\omega = 5$ μA and $\theta = 330°$. The $V_{xy}^{(2n+1)\omega}$ reaches its maximum value near the charge neutral point, indicating by the dashed line. **b,** Temperature-dependent higher odd-order nonlinear Hall voltage $V_{xy}^{(2n+1)\omega}$ at $I^\omega = 5$ μA and $\theta = 330°$ without applying gate voltages. The $V_{xy}^{(2n+1)\omega}$ exists only below the Néel temperature. **c,** Decay of higher odd-order nonlinear Hall voltage $V_{xy}^{(2n+1)\omega}$ at $T = 5$ K, $I^\omega = 5$ μA and $\theta = 330°$. The vertical coordinate is in exponential form. The $V_{xy}^{(2n+1)\omega}$ shows an exponential decay as the nonlinear order $N$ increases.

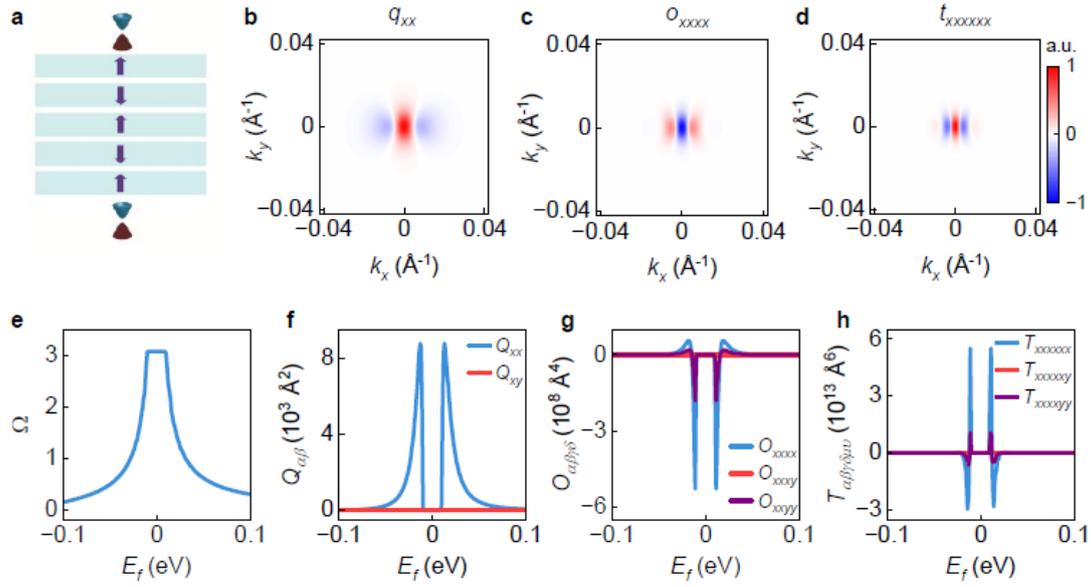

**Fig. 4 | Theoretical calculations of higher odd-order nonlinear Hall effect from Dirac cone surface states.**

**a,** Schematic of the top and bottom Dirac cone surface states of the 5SL Mn(Bi$_{1-x}$Sb$_x$)$_2$Te$_4$, with identical TRS-breaking Dirac gaps and resulting Berry curvatures induced by parallel magnetic ordering in the top and bottom surfaces. **b-d,** The $k$-resolved distributions for the dominant components of the higher odd-order Berry curvature multipoles, namely, the quadrupole component $q_{xx}$ (b), octapole component $o_{xxxx}$ (c), and dodecapole component $t_{xxxxxx}$ (d). **e-h,** The calculated Berry curvature monopole (e), quadrupole (f), octapole (g), and dodecapole (h) as a function of the Fermi level for each Dirac surface state at $T$ = 5 K, where peaks appear around the CNP.